\documentclass[twocolumn,twocolappendix,trackchanges]{aastex7}
\usepackage{bm}
\usepackage{braket}
\usepackage{amsmath}
\usepackage{subeqnarray}
\usepackage{here}
\usepackage{multirow}
\usepackage{makecell}
\usepackage{tabularx}
\usepackage{threeparttable}
\usepackage{lipsum}
\usepackage{lineno}
\usepackage[normalem]{ulem}

\received{June 15, 2026}
\revised{July 31, 2026}
\accepted{August 6, 2026}
\submitjournal{ApJ}
\shorttitle{}
\shortauthors{Yoshihisa $\&$ Yokoyama}
\begin{document}
\title{Coronal Rain Driven by Low-Frequency Impulsive Heating: Dependence of Dynamics and Morphology on Coronal Heating and Magnetic Field}

\correspondingauthor{Takero Yoshihisa}

\author[0000-0001-5457-4999]{Takero Yoshihisa}
\email{yoshihisa@kusastro.kyoto-u.ac.jp}
\affiliation{\rm{Department of Astronomy, Kyoto University, Sakyo-ku, Kyoto, 606-8502, Japan}}

\author[0000-0001-5457-4999]{Takaaki Yokoyama}
\email{yokoyama@kwasan.kyoto-u.ac.jp}
\affiliation{\rm{Astronomical Observatory, Kyoto University, Sakyo-ku, Kyoto 606-8502, Japan}}

\begin{abstract}

We performed numerical simulations to investigate how coronal heating and magnetic field strength regulate the dynamics and morphology of coronal rain in the solar corona.
Coronal rain is widely interpreted as a manifestation of thermal non-equilibrium and thermal instability. 
Many previous studies have investigated condensation under (quasi-)steady heating scenarios, but such heating models do not fully explain the observed properties, including multiple clumps and symmetric drainage toward both loop footpoints.
Condensation driven by low-frequency impulsive heating has been much less explored and may produce distinct coronal rain morphology and drainage dynamics.
The role of the coronal magnetic field also remains poorly understood.
We conducted 1.5-dimensional magnetohydrodynamic simulations along a single strand with self-consistent coronal heating, upon which a single impulsive heating event was imposed.
Two key parameters were systematically explored: the duration of localized heating and the strength of the coronal background magnetic field.
We find that condensation can be triggered even by a single impulsive heating event, provided that the injected energy is about an order of magnitude larger than that in the steady heating case.
Condensation becomes more likely for longer heating durations and weaker magnetic fields.
The dynamics of coronal rain depend strongly on the heating duration.
When the heating duration is shorter than the radiative cooling timescale, condensations drain toward both loop footpoints.
We also find that weaker magnetic fields enhance the nonlinearity of Alfvén waves, promoting fragmented condensations.
These results demonstrate that the heating duration and coronal magnetic field strength play key roles in regulating coronal rain formation and dynamics.

\end{abstract}


\section{Introduction} \label{sec:intro}

Coronal rain is cool and dense plasma condensations with temperatures of order $10^{4}-10^{5}$ K that form in the corona and subsequently fall toward the solar surface along magnetic field lines.
Observed in transition region and chromospheric spectral lines \citep{schrijver2001catastrophic,antolin2012observing}, coronal rain provides one of the most direct probes of the thermodynamic evolution of coronal loops \citep{antolin2010coronal}.

Observations reveal that coronal rain exhibits substantial morphological and dynamical complexity.
Individual rain blobs typically have transverse widths of about 120 km in H${\mathrm{\alpha}}$ and up to several hundred kilometers in transition-region lines, while their lengths range from 1–3 Mm to several tens of Mm due to stretching during descent.
Multiple condensations are often observed along similar trajectories, and rain structures frequently exhibit internal density inhomogeneities along magnetic strands \citep{antolin2015multi, antolin2012observing, vashalomidze2015formation}.
Both asymmetric drainage toward a single footpoint and symmetric drainage toward both loop footpoints have also been reported \citep{froment2020multi, csahin2023spatial}.
These observations suggest that the morphology and dynamics of coronal rain are diverse and non-uniform from one coronal loop to another.

Coronal rain is widely interpreted as a manifestation of thermal non-equilibrium (TNE) and thermal instability (TI) \citep{antolin2022multi,keppens2025modeling}.
In this framework, localized heating near loop footpoints drives chromospheric evaporation, increasing the coronal density 
$n$.
The enhanced density strengthens radiative cooling proportional to $n^2 \Lambda(T)$, where $\Lambda$ is the optically-thin cooling function, leading to runaway cooling and the formation of cool condensations \citep[e.g.,][]{mok1990prominence,antiochos1991model,antiochos1999dynamic}.
\citet{yang2021formation} reported observational evidence of chromospheric evaporation followed by coronal condensation, providing support for the evaporation--condensation scenario of filament formation.
\cite{csahin2022prevalence} showed that about half of active-region loop systems reside in TNE states.
The occurrence of such condensations depends sensitively on both the temporal and spatial properties of coronal heating.

Previous theoretical and numerical studies have demonstrated that high-frequency or quasi-steady footpoint heating can trigger condensation by maintaining high coronal densities during the cooling phase \citep{karpen2008condensation, johnston2019effects,kohutova2020self}.
Here, high-frequency or quasi-steady heating refers to heating in which the interval between heating events is shorter than the radiative cooling timescale ($\tau_{\mathrm{int}}<\tau_{\mathrm{rad}}$).
Sufficiently strong footpoint-localized heating relative to background heating is also a key condition for condensation onset \citep{klimchuk2019role}.
Simulations adopting such heating prescriptions have successfully reproduced long-period intensity pulsations and coronal rain formation \citep{mikic2013importance,fang2015coronal,froment2017long, pelouze2022role,lu2024periodic}, consistent with observations of long-period intensity pulsations \citep{auchere2014long,froment2015evidence,auchere2018coronal} and with independent evidence for quasi-steady, footpoint-concentrated heating in coronal loops \citep{aschwanden2000evidence,aschwanden2001modeling,ishigami2024spectroscopic}.
In these heating scenarios, evaporated plasma is continuously supplied to preferred locations along a coronal loop, leading to monotonic condensation and drainage patterns.
Such scenarios do not fully explain the observed multiple clumps or symmetric drainage toward both loop footpoints.

Low-frequency ($\tau_{\mathrm{int}} \gg \tau_{\mathrm{rad}}$) impulsive heating events may provide an alternative explanation for such observed features.
Here, impulsive heating refers to heating whose duration is much shorter than the radiative cooling timescale ($\tau_{\mathrm{heat}} \ll \tau_{\mathrm{rad}}$).
Under such heating conditions, cooling and condensation evolve after the localized heating has ceased, potentially producing condensation and drainage dynamics distinct from those in quasi-steady heating scenarios.
The subsequent thermodynamic evolution may therefore depend more sensitively on the background coronal environment.
Coronal heating is believed to contain both steady and impulsive components, as suggested by observations \citep{shimizu1995energetics} and numerical modeling \citep{breu2022solar, kuniyoshi2025unified, johnston2025self, sow2026relationship}.
Recent detections of nanoflare-like events, such as campfires in quiet-Sun regions, provide additional evidence for intermittent energy release \citep{berghmans2021extreme,alipour2022automatic}.
These results and observations support the importance of considering low-frequency impulsive heating in coronal rain formation.


In this study, we investigate the formation and evolution of condensations in semicircular coronal loops subjected to a single impulsive localized heating event.
To model the coronal loop, we employ one-and-a-half-dimensional (1.5D) magnetohydrodynamic (MHD) simulations that incorporate footpoint velocity perturbations and phenomenological turbulent heating term, enabling realistic coronal heating and atmospheric dynamics.
This framework allow us to examine how low-frequency impulsive heating influences the morphology and drainage dynamics of coronal rain.
We also systematically investigate the relationship between coronal magnetic fields and coronal rain, which has not yet been sufficiently clarified either observationally or theoretically.

We describe the numerical setting in Section~\ref{sec:numerical setup}. Section~\ref{sec:results} presents the simulation results and parameter survey. In Section~\ref{sec:Discussion}, we discuss the implication of these results. Our conclusion is summarized in Section \ref{sec:summary}.

\section{NUMERICAL SETUP}\label{sec:numerical setup}

The basic numerical setup is largely the same as that described in \cite{yoshihisa2025conditions}.
The main differences are limited to the loop geometry.
Therefore, we summarize only the essential parts of the model below.

\subsection{Coordination and Geometry}\label{geometry}

We consider a single coronal loop by an expanding flux tube rooted on the solar surface.
The coordinate `$s$' denotes the position along the loop, while `$x$' and `$y$' represent directions perpendicular to the loop.
$\bm{e}_{x}$ and $\bm{e}_{y}$ represent unit vectors along $x$- and $y$-axis, respectively.

The loop is assumed to be symmetric with respect to the apex at $s = L/2$, where $L=100$ Mm is the loop length.

The gravitational acceleration along the loop, $g(s)$, is given by
\begin{equation}\label{eq:gravity}
    g(s) = g_{\odot} \cos{\Bigl(\frac{\pi s}{L}\Bigr)},
\end{equation}
where $g_{\odot}$ is the gravitational acceleration at the solar surface.

The expansion factor, $f_{\mathrm{ex}}(s)$, describes the relative variation of the cross-sectional area of the loop, normalized to unity at the surface. It is given by
\begin{align}
    f_{\mathrm{ex}}(s) &= \left(\frac{1}{f_{\mathrm{ex,0}}(\tilde{s})^{10}} + \frac{1}{f_{\mathrm{ex,max}}^{10}} \right)^{-1/10},\label{eq:expansion factor2} \\
    f_{\mathrm{ex,0}}(\tilde{s}) &= \min \left[\exp \left(\frac{\mu g_{\odot} m_{\mathrm{H}}}{2 k_{\mathrm{B}} T_{\mathrm{surf}}} \tilde{s} \right), f_{\mathrm{ex,max}} \right], \label{eq:expansion factor}
\end{align}
where $f_{\mathrm{ex,max}}$ represents the maximum value of the expansion factor over $s$.
$\mu$, $k_{\mathrm{B}}$, $m_{\mathrm{H}}$, and $T_{\mathrm{surf}}$ are the constant variables listed in Table~\ref{tab:constant variables} and \ref{tab:boundary variables}.
$\tilde{s}$ represents the length along the loop from the closer footpoint given as,
\begin{equation}
    \tilde{s}(s) = \frac{L}{2} - \left|s - \frac{L}{2}\right|.
\end{equation}

\subsection{Basic Equations and Setting}\label{sec:basic equations and setting}

\begin{table}[t]
\centering
\caption{List of Main constant parameters.}
\begin{tabular}{lll}
\hline
 & Symbol & Value \\
\hline
Boltzmann constant & $k_{\mathrm{B}}$ & $1.38 \times 10^{-16} \ \mathrm{erg \ K^{-1}}$ \\
Hydrogen mass & $m_{\mathrm{H}}$ & $1.67\times 10^{-24} \ \mathrm{g}$ \\
Adiabatic index & $\gamma$ & 5/3 \\
Mean molecular weight & $\mu$  & 0.5 \\
Gravitational acc. & $g_{\odot}$ & $2.74\times 10^{4} \ \mathrm{cm \ s^{-1}}$ \\
Spitzer conductivity & $\kappa_{\mathrm{SH}}$ & $10^{-6} \ \mathrm{erg\ cm^{-1} \ s^{-1} \ K^{-7/2}}$ \\
Loop length & $L$ & 100 Mm \\
\hline
\end{tabular}
\label{tab:constant variables}
\end{table}

\begin{table}[t]
\centering
\caption{List of physical constants at the boundary.}
\begin{tabular}{lll}
\hline
 & Symbol & Value \\
\hline
Mass density & $\rho_{\mathrm{surf}}$ & $1.0 \times 10^{-7} \ \mathrm{g \ cm^{-3}}$ \\
Temperature & $T_{\mathrm{surf}}$ & $6000 \ \mathrm{K}$ \\
Magnetic Field strength & $B_{s, \mathrm{surf}}$ & $1000 \ \mathrm{G}$ \\
Longitudinal velocity & $v_{s, \mathrm{surf}}$ & $0.9 \ \mathrm{km \ s^{-1}}$ \\
Transverse velocity & $v_{x(y),  \mathrm{surf}}$ & $1.2 \ \mathrm{km \ s^{-1}}$ \\
Min. longitudinal Frequency & $f_{\min, s}$ & $1.0\times 10^{-3} \ \mathrm{Hz}$ \\
Max. longitudinal Frequency & $f_{\max, s}$ & $1.0\times 10^{-2} \ \mathrm{Hz}$ \\
Min. transverse Frequency & $f_{\min, x(y)}$ & $3.33\times 10^{-3} \ \mathrm{Hz}$ \\
Max. transverse Frequency & $f_{\max, x(y)}$ & $1.0\times 10^{-2} \ \mathrm{Hz}$ \\
Correlation length & $\lambda_{\mathrm{cor, surf}}$ & 100 km \\
\hline
\end{tabular}
\label{tab:boundary variables}
\end{table}

We solve the one-dimensional MHD equations including gravity, thermal conduction, radiative cooling (and heating), phenomenological turbulent dissipation, and footpoint-localized heating.
The governing equations are given by
\begin{gather}
    \frac{\partial}{\partial t}(\rho f_{\mathrm{ex}}) + \frac{\partial}{\partial s} (\rho v_s f_{\mathrm{ex}})=0, \label{mass} \\
    \frac{\partial}{\partial t} (\rho v_s f_{\mathrm{ex}}) + \frac{\partial}{\partial s} \left[\left(\rho v_{s}^{2} + p + \frac{\bm{B}_{\perp}^{2}}{8\pi}  \right) f_{\mathrm{ex}} \right] \nonumber \\
    = \left(p+\frac{\rho \bm{v}_{\perp}^{2}}{2} \right) \frac{d}{ds} f_{\mathrm{ex}} - \rho g f_{\mathrm{ex}}, \label{moment} \\
    \frac{\partial}{\partial t}(\rho \bm{v}_{\perp} f_{\mathrm{ex}}^{3/2}) + \frac{\partial}{\partial s}\left[\left(\rho v_s \bm{v}_{\perp} - \frac{B_s \bm{B}_{\perp}}{4\pi} \right)f_{\mathrm{ex}}^{3/2} \right] \nonumber \\
    = - \hat{\bm{\eta}}_{+}\cdot \rho \bm{v}_{\perp} f_{\mathrm{ex}}^{3/2} - \hat{\bm{\eta}}_{-} \cdot \sqrt{\frac{\rho}{4\pi}} \bm{B}_{\perp} f_{\mathrm{ex}}^{3/2}, \label{vperp} \\
    \frac{\partial}{\partial t} (\bm{B}_{\perp}\sqrt{f_{\mathrm{ex}}}) + \frac{\partial}{\partial s}[(\bm{B}_{\perp}v_{s} - B_s \bm{v}_{\perp})\sqrt{f_{\mathrm{ex}}}] \nonumber \\
    = - \hat{\bm{\eta}}_{+} \cdot \bm{B}_{\perp}\sqrt{f_{\mathrm{ex}}} - \hat{\bm{\eta}}_{-} \cdot \sqrt{4\pi\rho} \bm{v}_{\perp} \sqrt{f_{\mathrm{ex}}}, \label{Bperp} \\
    \frac{d}{ds}(f_{\mathrm{ex}} B_s) = 0, \label{divB} \\
    \frac{\partial}{\partial t}\left[\left(e + \frac{1}{2}\rho v^2 + \frac{\bm{B}^2}{8\pi} \right) f_{\mathrm{ex}} \right] \nonumber \\
    + \frac{\partial}{\partial s} \left[\left(e + p + \frac{1}{2}\rho v^2 + \frac{\bm{B}_{\perp}^{2}}{4\pi} \right) v_{s} f_{\mathrm{ex}} - B_{s} \frac{\bm{B}_{\perp}\cdot \bm{v}_{\perp}}{4 \pi}f_{\mathrm{ex}} \right] \nonumber \\
    = f_{\mathrm{ex}} (-\rho g v_{s} - Q_{\mathrm{rad}} + Q_{\mathrm{cond}} + Q_{\mathrm{L}}), \label{EnergyEq} \\
    e = \frac{p}{\gamma-1}, \label{energy} \\
    p = \frac{\rho k_{\mathrm{B}} T}{\mu m_{\mathrm{H}}}. \label{EoS}
\end{gather}
Here, $\rho$, $p$, $T$, $\bm{v}$, and $\bm{B}$ denote the mass density, gas pressure, temperature, velocity, and magnetic field, respectively, using standard notation in MHD.
$\bm{v}_{\perp} = v_{x}\bm{e}_{x}+v_{y}\bm{e}_{y} $ and $\bm{B}_{\perp} = B_{x}\bm{e}_{x} + B_{y}\bm{e}_{y} $ are the transverse components of velocity and magnetic fields, respectively.
The values of constant parameters are summarized in Table~\ref{tab:constant variables}.

The magnetic field strength along the loop, $B_{s}(s)$, is given by
\begin{gather}
    B_{s}(s) = \frac{1}{f_{\mathrm{ex}}(s)} B_{s,\mathrm{surf}}.
\end{gather}
The strength of the coronal magnetic field is defined by $B_{s,\mathrm{cor}}=B_{s,\mathrm{surf}}/f_{\mathrm{ex, max}}$.

The formulation of the turbulent dissipation follows \cite{shoda2018frequency,shoda2018self} and \cite{shoda2021corona}.
$\hat{\eta}_{\pm}$, coefficient tensors in Equations (\ref{vperp}) and (\ref{Bperp}), are implemented like below:
\begin{gather}
    \hat{\bm{\eta}}_{\pm} = \frac{c_{d}}{4\lambda_{\mathrm{cor}}}[(|\xi_{x}^{+}| \pm |\xi_{x}^{-}|)\bm{e}_{x}\bm{e}_{x} + (|\xi_{y}^{+}|\pm|\xi_{y}^{-}|)\bm{e}_{y}\bm{e}_{y}], \label{eta1} 
\end{gather}
where $\xi_{x}^{\pm}$ and $\xi_{y}^{\pm}$ are Elsasser variables \citep{elsasser1950hydromagnetic}:
\begin{gather}
    \xi_{x,y}^{\pm} = v_{x,y} \mp \frac{B_{x,y}}{\sqrt{4\pi \rho}}.
\end{gather}
$\lambda_{\mathrm{cor}}$ is the correlation length perpendicular to the mean field. We assume it increases with the expansion of the flux tube:
\begin{equation}
    \lambda_{\mathrm{cor}} = \lambda_{\mathrm{cor,surf}} \sqrt{\frac{B_{s,\mathrm{surf}}}{B_{s}}},
\end{equation}
where $\lambda_{\mathrm{cor,surf}}$ is a typical length of inter granular lanes.
We adopt $c_{d}=0.25$.
The dissipation terms associated with $\hat{\bm{\eta}}_{\pm}$ reduce the kinetic and magnetic energies of the transverse fluctuations.
Since the total energy,
\begin{equation}
    E_{\mathrm{tot}} = e + \frac{1}{2}\rho v^{2} + \frac{B^{2}}{8\pi},
\end{equation}
is evolved without introducing a corresponding sink term for the turbulent dissipation, the energy removed from the transverse fluctuations is retained in the total energy and is consequently converted into internal energy.
Thus, the heating associated with Alfv\'en wave turbulent dissipation is included implicitly through the damping terms in Equations~(\ref{vperp}) and~(\ref{Bperp}).

The footpoint-localized heating rate, $Q_{\mathrm{L}}(s,t)$, is introduced to trigger chromospheric evaporation. It is prescribed as a Gaussian function in space and time, written as
\begin{gather}
    Q_{\mathrm{L}}(s,t) = Q_{\mathrm{L, peak}} F(s)G(t) \label{eq:QL} \\
    F(s) = \exp{\left( - \frac{(s-s_{\mathrm{peak}})^{2}}{\ell_{\mathrm{w}}^{2}} \right)} \label{eq:Localized heating space AB} \\
    G(t) = \exp{\left(- \frac{(t-t_{\mathrm{peak}})^{2}}{\tau_{\mathrm{w}}^{2}} \right)} \label{eq:Localized heating time AB},
\end{gather}
where $Q_{\mathrm{L, peak}}$ is the maximum heating rate and $s_{\mathrm{peak}}$ and $t_{\mathrm{peak}}$ denote the location and time at which the term peaks.
The parameters $\ell_{\mathrm{w}}$ and $\tau_{\mathrm{w}}$ represent the spatial and temporal widths of the Gaussian functions, respectively.

To realistically reproduce the solar atmosphere, both optically thin and thick radiative cooling functions are included in the radiative cooling term $Q_{\mathrm{rad}}$.
The optically thick cooling approximates radiative losses in the lower atmosphere following \citet{gudiksen2005ab}.
The optically thin cooling consists of chromospheric and coronal components.
The chromospheric cooling is described by \citet{goodman2012radiating}, whereas the coronal cooling is calculated using the Chianti Atomic Database ver. 7.0 \citep{dere1997chianti,landi2011chianti}.
For the thermal conduction term $Q_{\mathrm{cond}}$, we employ the Spitzer--H\"arm conduction \citep{spitzer1953transport}.
Details of the radiative cooling and thermal conduction formulations are described in \citet{yoshihisa2025conditions}.

\subsection{Initial and Boundary Conditions}\label{sec:initial condition}

The initial atmosphere is set to be static ($v=0$) and uniform temperature with $T=T_{\mathrm{surf}}$ along the loop.
This uniformly cool initial condition is adopted to examine whether a hot coronal atmosphere can be produced and maintained self-consistently by the wave-driven heating included in our model.
The density and pressure are stratified gravitationally, with $\rho=\rho_{\mathrm{surf}}$ and $p=p_{\mathrm{surf}}$ at the surface.

At the boundary, the density and temperature are fixed at $\rho=\rho_{\mathrm{surf}}$ and $T=T_{\mathrm{surf}}$. Velocity perturbations with a pink-noise spectrum are injected both parallel and perpendicular to the loop to excite Alfv\'en and acoustic waves.
The frequency ranges are $f_{\min, s}$ to $f_{\max, s}$ in the longitudinal direction and $f_{\min, x(y)}$ to $f_{\max, x(y)}$ in the transverse direction.
The root-mean-square amplitudes of the velocity perturbation are $v_{s,\mathrm{surf}}$ and $v_{x(y),\mathrm{surf}}$ for the longitudinal and transverse components, consistent with the observational values.
The values of corresponding parameters are summarized in Table \ref{tab:boundary variables}.

Once the loop reaches a quasi-steady state, the localized heating term, $Q_{\mathrm{L}}$, is introduced.
The injection of waves from the boundary is maintained throughout the simulation.

\subsection{Parameter Survey Setup}

To investigate the dynamics and morphology of coronal rain, we systematically vary two parameters: the localized heating duration $\tau_{\mathrm{w}}$ and the background coronal magnetic field strength $B_{s,\mathrm{cor}}$.

For localized heating, we define an effective heating duration, $\tau_{\mathrm{w,eff}}$, as the time interval during which the localized heating rate exceeds the background heating rate.
Here, the background heating refers to the heating produced by the
dissipation of MHD waves injected from the lower boundaries through
turbulent and shock dissipation.
The value of $\tau_{\mathrm{w,eff}}$ ranges from approximately 3 s to 5400 s, covering both nanoflare-like impulsive heating inferred from observations \citep{berghmans2021extreme, patel2022hi, alipour2022automatic, wallace2025reconnection} and quasi-steady heating exceeding the radiative cooling timescale ($\tau_{\mathrm{w,eff}}>\tau_{\mathrm{rad}}$) adopted in previous studies \citep[e.g.,][]{xia2011formation}.

The background coronal magnetic field strength $B_{s,\mathrm{cor}}$ is controlled by varying the maximum expansion factor $f_{\mathrm{ex,max}}$ and spans a range of $1$--$100 \ \mathrm{G}$.

The total injected energy is calculated as
\begin{equation*}
    E_{\mathrm{inj}} = \int_{0}^{L/2} ds \int_{t_{\mathrm{ini}}}^{t_{\mathrm{end}}}dt  \ Q_{\mathrm{L}}(s,t).
\end{equation*}
In all cases, we consider only the energy injected into the left half
of the loop, corresponding to $0 \leq s \leq L/2$.
The value of $E_{\mathrm{inj}}$ is controlled by adjusting the peak heating rate $Q_{\mathrm{L,peak}}$ and $\tau_{\mathrm{w}}$ in Equations~\eqref{eq:QL} and ~\eqref{eq:Localized heating time AB}, and ranges from approximately $2.0 \times 10^{10}$ to $9.5 \times 10^{11}\ \mathrm{erg\,cm^{-2}}$.

\cite{yoshihisa2025conditions} showed that one-sided localized heating requires several times more energy input to trigger condensation than both-sided localized heating.
Therefore, one-sided heating is more prone to numerical instability.
While nanoflare-like heating is expected to occur stochastically at localized positions along coronal loops, we impose localized heating symmetrically at both loop footpoints in most simulations for simplicity.
To evaluate the impact of this simplification, we also perform a limited number of simulations with one-sided heating.

\begin{figure}[!]
  \epsscale{1.0}
  \plotone{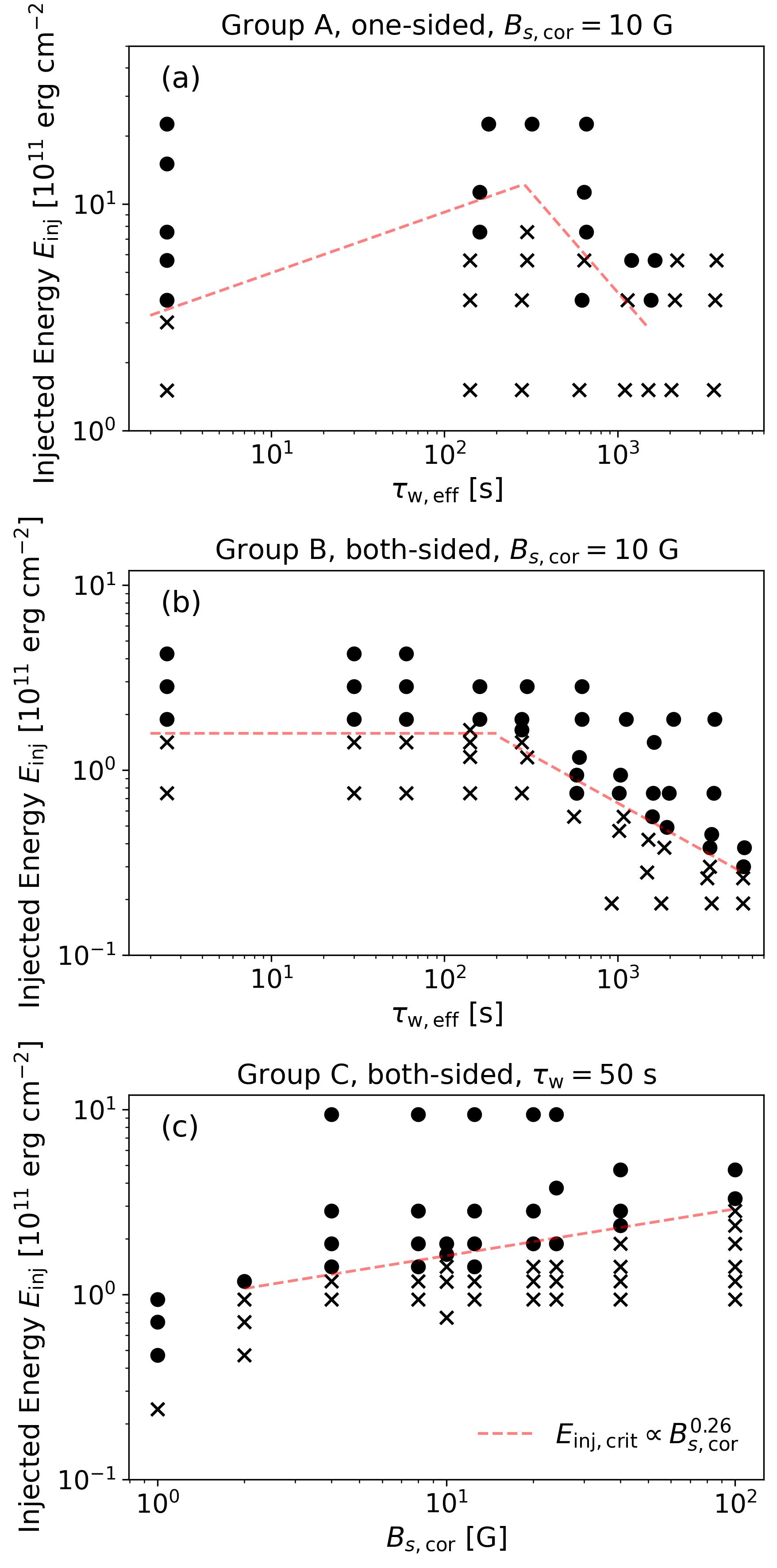}
  \caption{Parameter space of the simulations showing whether condensation occurs.
  The simulations are categorized into three groups: Group A corresponds to one-sided heating cases with varying $\tau_{\mathrm{w}}$ (Panel (a)), Group B corresponds to both-sided (symmetry) heating cases with varying $\tau_{\mathrm{w}}$ (Panel (b)), and Group C corresponds to both-sided heating cases with varying $B_{s,\mathrm{cor}}$ (Panel (c)).
  The horizontal axis represents the effective heating duration $\tau_{\mathrm{w,eff}}$ in Panels (a) and (b), and $B_{s,\mathrm{cor}}$ in Panel (c).
  The vertical axis shows the injected energy (in $10^{11} \ \mathrm{erg \ cm^{-2}}$) in all panels. 
  Circles and crosses denote cases with and without condensation, respectively. The red dashed lines indicate approximate boundaries for condensation, corresponding to the critical injected energy, $E_{\mathrm{inj,crit}}$. }
  \label{fig:parameter survey}
\end{figure}

The simulations are categorized into three groups.
Group A consists of one-sided heating cases with varying $\tau_{\mathrm{w}}$.
Group B consists of both-sided (symmetry) heating cases with varying $\tau_{\mathrm{w}}$. 
Group C consists of both-sided heating cases with varying $B_{s,\mathrm{cor}}$.
In Group A and B, $B_{s, \mathrm{cor}}$ is fixed at $10 \ \mathrm{G}$, and in Groups C, $\tau_{\mathrm{w,eff}}$ is fixed at $300 \ \mathrm{s}$.
The parameter space is summarized in Figure~\ref{fig:parameter survey}.

\subsection{Numerical Scheme}

We solve the MHD equations using the fourth-order central finite difference scheme with artificial viscosity in space and the fourth-order Runge–Kutta method in time \citep{vogler2005simulations,rempel2009radiative}. Thermal conduction is calculated simultaneously using the super-time-stepping method, which is second-order in both space and time \citep{meyer2012second,meyer2014stabilized}.
We use a uniform grid size of $10 \ \mathrm{km}$.

\begin{figure}[!]
  \epsscale{1.0}
  \plotone{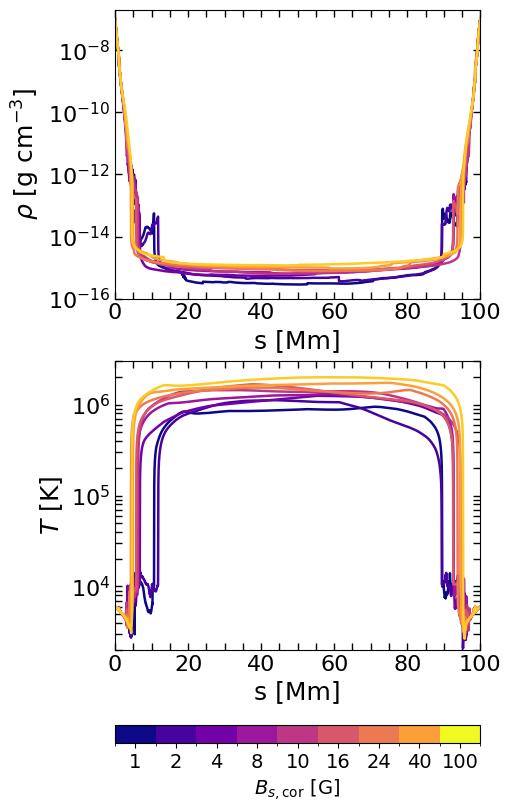}
  \caption{Density (top panel) and temperature (bottom panel) profiles at 30,000 s after the initial condition when the system reached a quasi-steady state. Different colors indicate cases with $B_{s,\mathrm{cor}}=1, \ 2, \ 4, \ 8, \ 10, \ 16, \ 24, \ 40, \ 100 \ \mathrm{G}$.}
  \label{fig:IC}
\end{figure}

\section{Results}\label{sec:results}

\subsection{Formation of Coronal Loops}\label{sec:Coronal Loops}

We first reproduce a quasi-steady coronal loop sustained by energy injection from velocity perturbations at the surface.
Figure~\ref{fig:IC} shows the density and temperature profiles of the coronal loop in a quasi-steady state obtained after $30{,}000\ \mathrm{s}$ from the initial condition described in Section~\ref{sec:initial condition}.
Different colors correspond to cases with different coronal magnetic field strengths.

\begin{figure}[!]
  \epsscale{1.2}
  \plotone{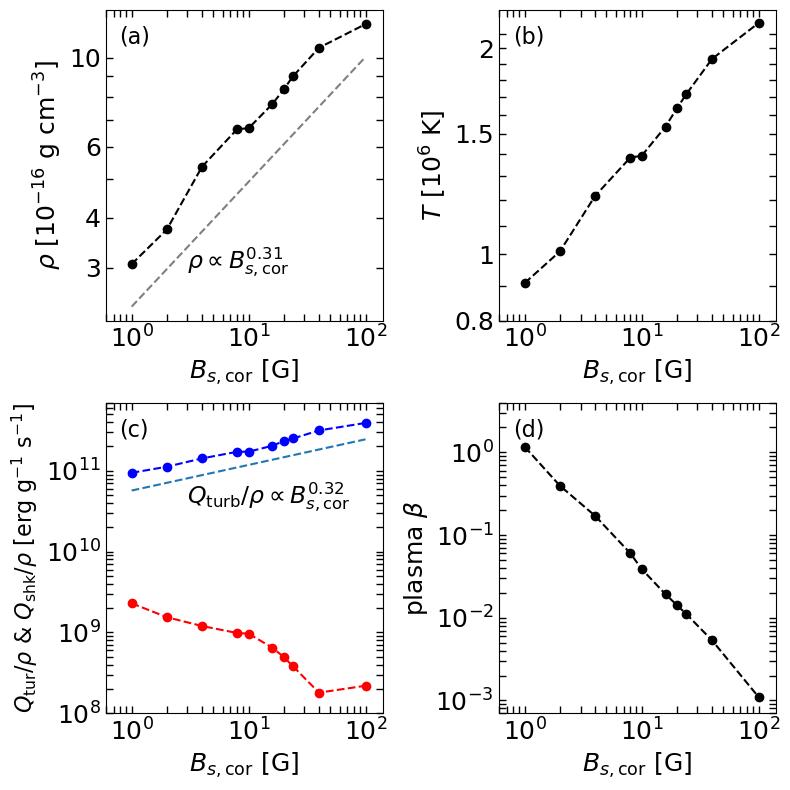}
  \caption{Density (top left), temperature (top right), turbulent (blue) and shock (red) heating rates (bottom left), and plasma $\beta$ (bottom right) at the loop apex are plotted.}
  \label{fig:Bs_vs_roteQtbeta}
\end{figure}

Figure~\ref{fig:Bs_vs_roteQtbeta} shows that the background coronal properties depend on the coronal magnetic field strength.
We adopt the $B_{s,\mathrm{cor}} = 10\ \mathrm{G}$ case as the representative one for the coronal rain simulations throughout this study.
The loop apex in this case attains typical coronal conditions with $\rho \sim 7 \times 10^{-16}\ \mathrm{g\ cm^{-3}}$ and $T \sim 1.4\ \mathrm{MK}$.
The turbulent heating rate is given by \citep{cranmer2007self,verdini2007alfven}
\begin{equation}
\label{Qturb}
    Q_{\mathrm{turb}} = \frac{c_{d}}{4\lambda_{\mathrm{cor}}} \rho \sum_{i=x,y} (|\xi_{i}^{+}|{\xi_{i}^{-}}^{2} + |\xi_{i}^{-}|{\xi_{i}^{+}}^{2}).
\end{equation}
The shock heating rate is estimated following \cite{shoda2018self} as
\begin{gather}
    \overline{Q_{\mathrm{shock}}} = \overline{ \frac{\partial e}{\partial t} } + \overline{ v_{s} \frac{\partial e}{\partial s} }
    + \overline{ \frac{e+p}{f_{\mathrm{ex}}} \frac{\partial}{\partial s} (v_{s} f_{\mathrm{ex}}) } \\
    - \overline{ Q_{\mathrm{cond}} } + \overline{ Q_{\mathrm{rad}} } - \overline{ Q_{\mathrm{turb}} }. \nonumber
    \label{eq:Qshockave}
\end{gather}
A comparison of the turbulent and shock heating rates shows that the former exceeds the latter by more than an order of magnitude.
Therefore, in the following discussion, we treat turbulent heating
as the dominant component of the background heating.
Increasing $B_{s,\mathrm{cor}}$ leads to higher density, temperature, and turbulent heating rates, while reducing the plasma $\beta$.
Since the plasma $\beta$ significantly exceeds unity for $B_{s,\mathrm{cor}} < 1\ \mathrm{G}$, such cases are excluded from the present simulations.

In this paper, these results are used as the initial conditions for the coronal rain simulations, and a more detailed discussion of the underlying physics will be presented in a future study.

\subsection{Plasma Response to Footpoint-Localized Heating}\label{sec:coronal rain formation}

After a quasi-steady coronal loop is established as in Section~\ref{sec:Coronal Loops}, the localized heating is introduced.
In real sun, localized heating is physically expected to occur at specific locations within coronal loops, making the one-sided heating adopted in Group A more realistic.
\cite{yoshihisa2025conditions} showed that although the heating energy required for condensation differs between one-sided and both-sided heating, the underlying physical processes remain essentially unchanged.
Therefore, in the present study, we mainly adopt the computationally more stable both-sided heating configuration (Groups B and C) to perform a broad parameter survey efficiently and robustly.
In all cases, $s_{\mathrm{peak}}$ in Equation~\eqref{eq:Localized heating space AB} is fixed at 10 Mm.
For both-sided cases, the localized heating is applied at both loop footpoints symmetrically by replacing $s$ with $\tilde{s}$ in the right-hand side of Equation~\eqref{eq:Localized heating space AB}.

\subsubsection{A representative case}\label{sec:representative case}

\begin{figure*}[!]
  \epsscale{1.2}
  \plotone{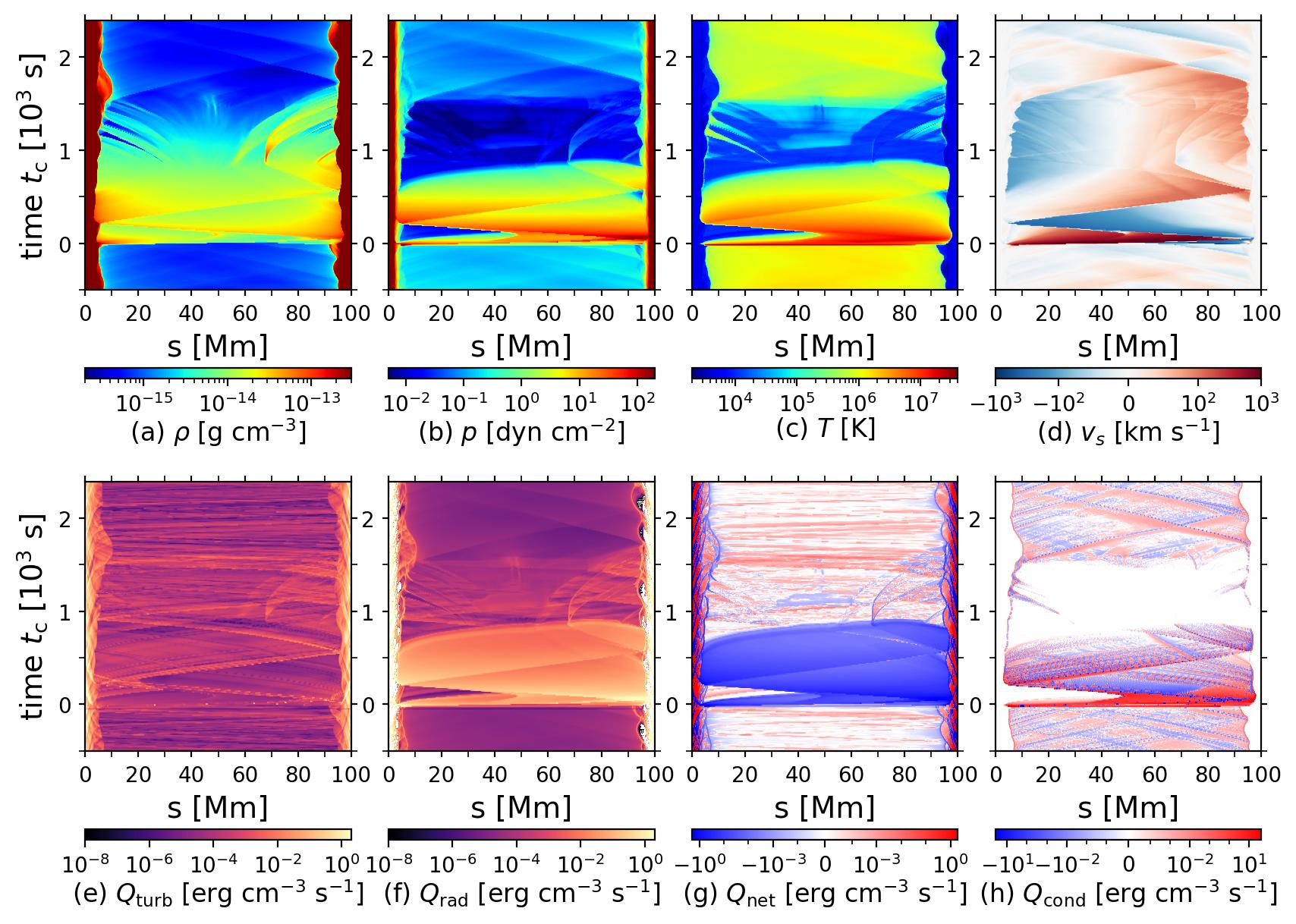}
  \caption{Space--time diagrams of a representative case in Group A (one-sided heating). Panels (a)-(h) show the density $\rho$, pressure $p$, temperature $T$, velocity along the loop $v_{s}$, turbulent heating rate $Q_{\mathrm{turb}}$, radiative cooling rate $Q_{\mathrm{rad}}$, net heating rate $Q_{\mathrm{net}}$, and thermal conduction $Q_{\mathrm{cond}}$ from $t_{\mathrm{c}}=-500$ to $2400 \ \mathrm{s}$. The horizontal and vertical axes indicate the length along the loop (in Mm) and time (in $10^3$ s). }
  \label{fig:roprtevs_QtQrQnQC}
\end{figure*}

Figure~\ref{fig:roprtevs_QtQrQnQC} shows the space--time diagrams of a representative case in Group A (one-sided heating, $B_{s,\mathrm{cor}}=10\ \mathrm{G}$, $\tau_{\mathrm{w,eff}} \sim 3\ \mathrm{s}$, and $E_{\mathrm{inj}} \sim 7.5\times 10^{11}\ \mathrm{erg\ cm^{-2}}$).
Hereafter, we use the time variable $t_{\mathrm{c}} = t - t_{\mathrm{peak}}$, measured from the peak of the localized heating.

Following the injection of localized heating at $t_{\mathrm{c}}\sim 0 \ \mathrm{s}$, thermal conduction transports heat flux (Figure~\ref{fig:roprtevs_QtQrQnQC}(h)), causing the coronal temperature to rise to $\sim 40$ MK (Figure~\ref{fig:roprtevs_QtQrQnQC}(c)).
The conducted heat flux also heats the transition region at the opposite footpoint, driving upflows from both footpoints despite the one-sided heating.
The resulting increase in coronal density strengthens radiative cooling and leads to a negative net heating rate $Q_{\mathrm{net}}$ ($=Q_{\mathrm{turb}}-Q_{\mathrm{rad}}$) (Figure~\ref{fig:roprtevs_QtQrQnQC}(a, f, g)).
At $t_{\mathrm{c}} = 500$--$900 \ \mathrm{s}$, the coronal temperature drops to $\sim 10^{4}$ K, and multiple cool plasma blobs condense and fall toward both footpoints along the loop.
The falling speed reaches approximately 360 km s$^{-1}$, which is faster than the free-fall speed.
This is attributed to the fact that the corona cools preferentially from the relatively dense footpoint regions, causing the pressure gradient to act in the same direction as gravity.
Such accelerated downflows may provide a possible explanation for the supersonic downflows observed in coronal rain \citep{kleint2014detection,ishikawa2020temporal,song2025two}.

At $t_{\mathrm{c}} = 20$--$230 \ \mathrm{s}$, low-pressure and low-temperature plasma appears on the left side of the corona.
This feature is caused by the expansion of the plasma accompanied by a shock wave following the localized energy injection.
The behavior is analogous to the Sedov--Taylor solution for a point-like explosion (\citealt{Sedov1959_ch4}, Chapter 4; \citealt{taylor1950formation}), while the present evolution is strongly modified by the dense lower coronal boundary.
Near the boundary, chromospheric evaporation increases the density and net radiative cooling, leading to decreases in temperature and pressure and the formation of a cool, low-pressure rarefaction region.


\subsubsection{Parameter survey overview}\label{sec:parameter survey}

Figure~\ref{fig:parameter survey} summarizes whether condensation occurs in each simulation group.
Panels (a) and (b) show, respectively, the results for Groups A and B, in which $\tau_{\mathrm{w,eff}}$ is varied.
In Group B, condensation occurs over the entire range $\tau_{\mathrm{w,eff}} \sim 3$–$5400$ s. 
In Group A, condensation does not occur for $\tau_{\mathrm{w,eff}} \gtrsim 2000$ s. 
This is because, in the one-sided heating cases with $\tau_{\mathrm{w,eff}} > \tau_{\mathrm{rad}}$, the heating asymmetry drives strong flows along the loop, which advect the cooled plasma and prevent the formation of condensation \citep[e.g.,][]{mikic2013importance,froment2018occurrence, klimchuk2019role}.

The dependence of the critical injected energy required for condensation  $E_{\mathrm{inj,crit}}$ on $\tau_{\mathrm{w,eff}}$ also differs between Groups A and B.
In Group B, $E_{\mathrm{inj,crit}}$ increases with decreasing $\tau_{\mathrm{w,eff}}$ and becomes nearly saturated for $\tau_{\mathrm{w,eff}} \lesssim 150$ s.
The saturation value is at most about an order of magnitude larger than that for the longest $\tau_{\mathrm{w,eff}}$.
In Group A, $E_{\mathrm{inj,crit}}$ also increases as $\tau_{\mathrm{w,eff}}$ decreases down to several hundred seconds.
For shorter durations, $E_{\mathrm{inj,crit}}$ decreases again because strong localized heating drives significant evaporation from the opposite chromosphere.

Panel (c) shows the results for Group C, in which the coronal magnetic field strength $B_{s,\mathrm{cor}}$ is varied. 
We find that $E_{\mathrm{inj,crit}}$ increases with increasing $B_{s,\mathrm{cor}}$.

\subsection{Dependence of Dynamics on Heating Duration}\label{sec:Dependence on Heating Duration}

\begin{figure}[!]
  \epsscale{1.0}
  \plotone{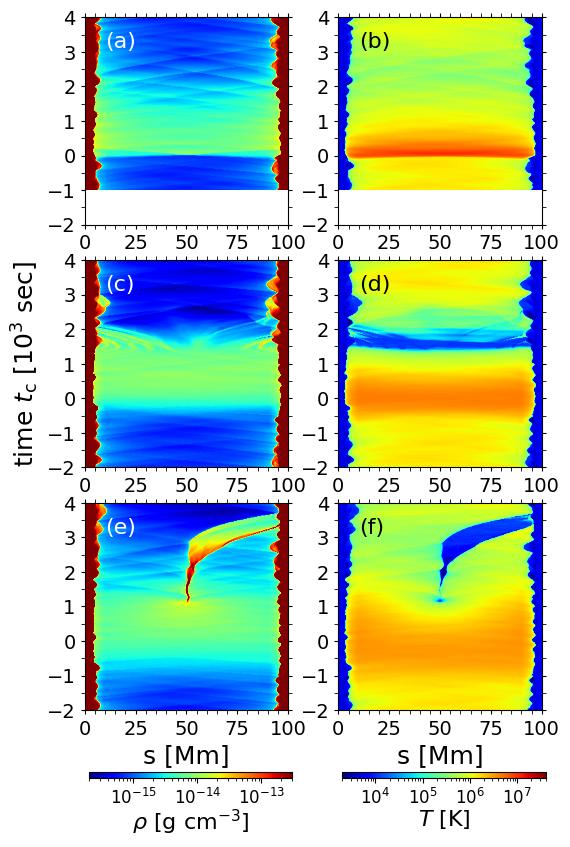}
  \caption{Space-–time diagrams of $\rho$ (left) and $T$ (right) for different heating durations $\tau_{\mathrm{w}}$ in Group B at a fixed injected energy, $E_{\mathrm{inj}} = 0.75\ \mathrm{erg\ cm^{-2}}$. 
  Panels (a) and (b), (c) and (d), and (e) and (f) correspond to $\tau_{\mathrm{w,eff}} = 300$, 2000, and 3600 s, respectively. 
  The color-range and axes are the same as Figure~\ref{fig:roprtevs_QtQrQnQC}. }
  \label{fig:compare tau_w}
\end{figure}

Figure~\ref{fig:compare tau_w} shows the results for Group B (both-sided heating). 
Here, the duration of localized heating is varied, while the injected energy ($E_{\mathrm{inj}} = 0.75\times10^{11}\ \mathrm{erg\ cm^{-2}}$) and the background coronal magnetic field ($B_{s,\mathrm{cor}}=10 \ \mathrm{G}$) are fixed.

As already shown in Figure~\ref{fig:parameter survey}, condensation does not occur for $\tau_{\mathrm{w,eff}} = 300\ \mathrm{s}$, whereas it does occur for $\tau_{\mathrm{w,eff}} = 2000$ and $3600\ \mathrm{s}$.
For $\tau_{\mathrm{w,eff}} = 2000\ \mathrm{s}$, the effective heating duration is shorter than the radiative cooling timescale ($\tau_{\mathrm{w,eff}} < \tau_{\mathrm{rad}} \sim 3000 \ \mathrm{s}$).
Similar to the representative case shown in Section~\ref{sec:representative case}, the entire corona cools down to chromospheric temperatures, after which multiple condensations form and fall toward both footpoints as coronal rain.
For $\tau_{\mathrm{w,eff}} = 3600\ \mathrm{s}$, the effective heating duration exceeds the radiative cooling timescale ($\tau_{\mathrm{w,eff}} > \tau_{\mathrm{rad}}$).
Since the evaporation flows from both footpoints are symmetric, plasma accumulates at the apex of the loop, resulting in the formation of a localized cool region with a length of $\sim 1\ \mathrm{Mm}$.
Subsequently, a single condensation falls toward one footpoint.
This behavior is consistent with that found in coronal rain under steady heating in previous studies \citep[e.g.,][]{mikic2013importance}.

\begin{figure}[!]
  \epsscale{1.0}
  \plotone{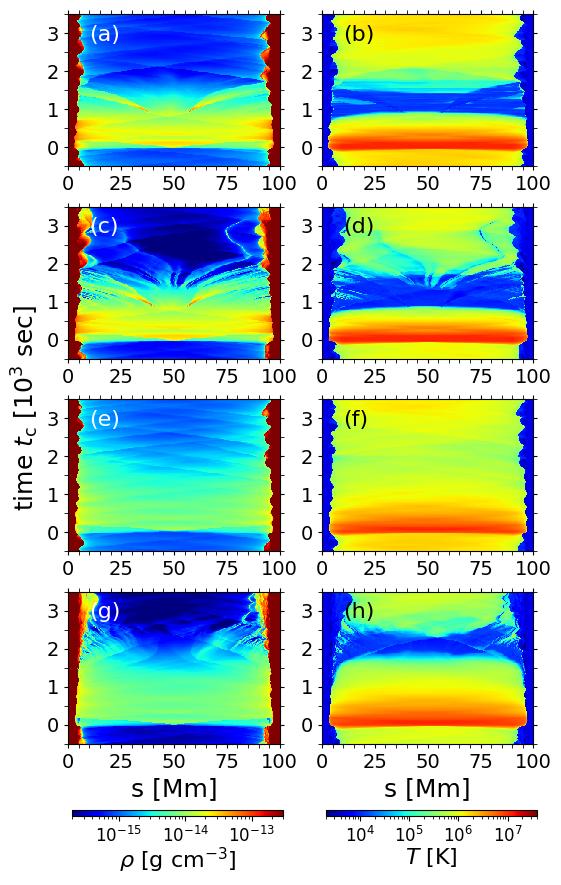}
  \caption{Same as Figure~\ref{fig:compare tau_w}, but for different coronal magnetic field strengths in Group C. 
  Panels (a) and (b), (c) and (d), (e) and (f), and (g) and (h) correspond to $B_{s,\mathrm{cor}} = 24,\ 4,\ 24,\ 1\ \mathrm{G}$. Panels (a)--(d) show cases with $E_{\mathrm{inj}} = 9.42\ \mathrm{erg\ cm^{-2}}$, while panels (e)--(h) show cases with $E_{\mathrm{inj}} = 0.94\ \mathrm{erg\ cm^{-2}}$.}
  \label{fig:compare Bs}
\end{figure}

\subsection{Dependence of Morphology on Coronal Magnetic Field}\label{sec:Dependence on Magnetic Field Strength}

Figure~\ref{fig:compare Bs} shows the results for Group C.
Except for the case with $B_{s,\mathrm{cor}}=24$ G and $E_{\mathrm{inj}}=0.94 \ \mathrm{erg\ cm^{-2}}$ (Fig.~\ref{fig:compare Bs} Panel (e, f)), the entire corona cools, and coronal rain falls toward both footpoints.
This behavior is consistent with the cases in which the heating duration is shorter than the radiative cooling timescale ($\tau_{\mathrm{w,eff}} < \tau_{\mathrm{rad}}$), as discussed in Section~\ref{sec:Dependence on Heating Duration}.

It is also found that the morphology of coronal rain depends on the magnetic field strength.
In the stronger-field case ($B_{s,\mathrm{cor}} = 24\ \mathrm{G}$, Fig.~\ref{fig:compare Bs}~(a, b)), a small number of condensations form and are stretched from $\sim 1$ to $\sim 10 \ \mathrm{Mm}$ during their descent. 
This stretching occurs because the effective gravitational acceleration differs between the upper and lower parts of each condensation.
In the weaker-field cases ($B_{s,\mathrm{cor}} = 4$ and $1\ \mathrm{G}$, Fig.~\ref{fig:compare Bs}~(c, d) and (g, h)), many condensations form over the corona, especially near the coronal base and fall with lengths of $\sim 1$--$2\ \mathrm{Mm}$.
Such footpoint condensations become more prominent toward weaker magnetic fields, particularly for $B_{s,\mathrm{cor}} \lesssim 8\ \mathrm{G}$.

\section{Discussion} \label{sec:Discussion}  

\subsection{Condition for Condensation}\label{sec:Discussion condensation condition}

\begin{figure}[!]
  \epsscale{1.0}
  \plotone{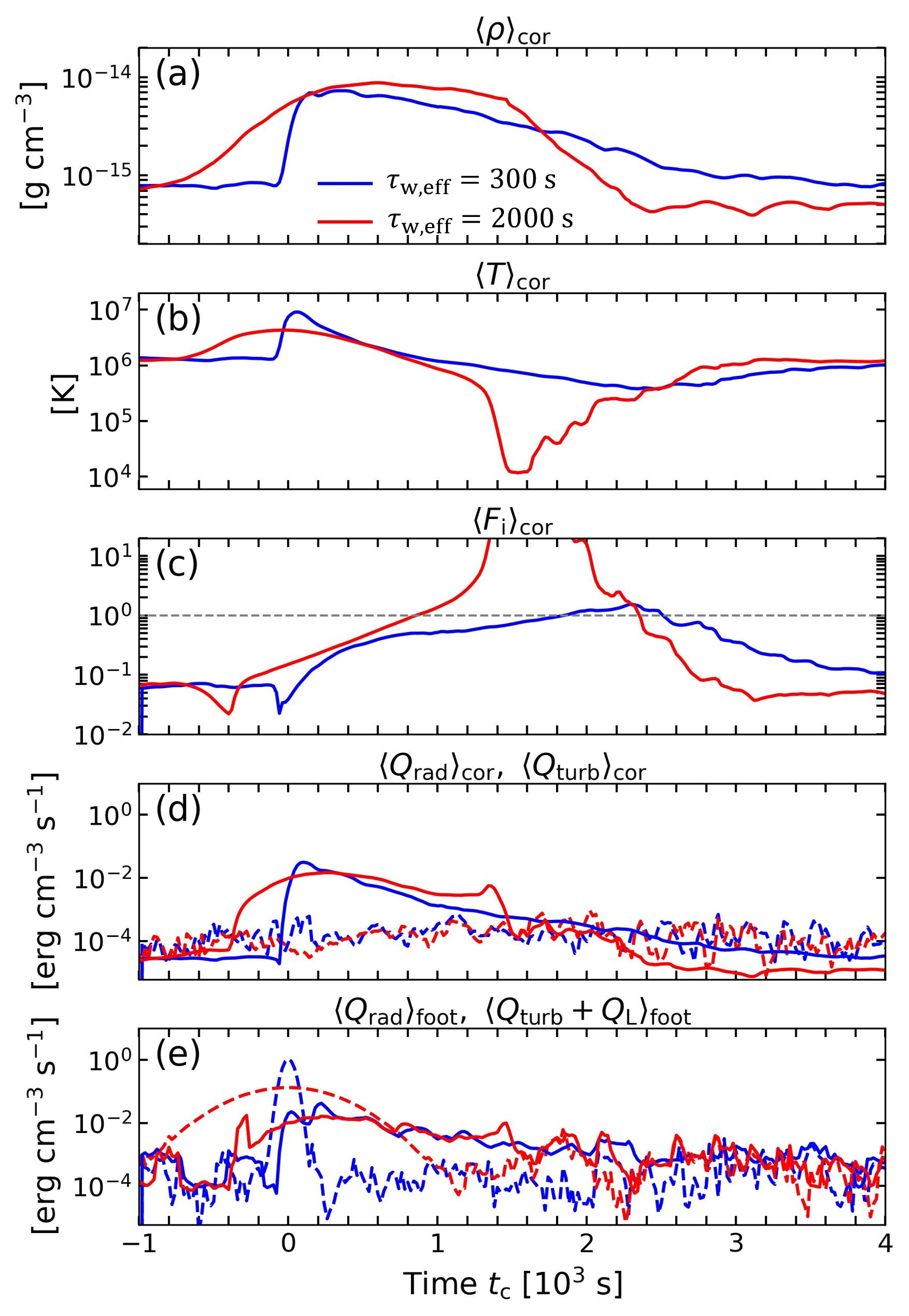}
  \caption{Time evolutions of various physical quantities for two cases in Group B with the same injected energy, $E_{\mathrm{inj}}=0.75 \times 10^{11} \ \mathrm{erg \ cm^{-2}}$, are shown.
  The blue and red lines indicate the case without ($\tau_{\mathrm{w,eff}}=300\ \mathrm{s}$) and with ($\tau_{\mathrm{w,eff}}=2000\ \mathrm{s}$) condensation, respectively.
  Panels (a)--(d) show coronal-averaged density $\braket{\rho}_{\mathrm{cor}}$, temperature $\braket{T}_{\mathrm{cor}}$, Field number $\braket{F_{\mathrm{i}}}_{\mathrm{cor}}$, and cooling rate $\braket{Q_{\mathrm{rad}}}_{\mathrm{cor}}$ (solid line) together with the turbulent heating rate $\braket{Q_{\mathrm{turb}}}_{\mathrm{cor}}$ (dashed line).
  Panel (e) shows the cooling rate $\braket{Q_{\mathrm{rad}}}_{\mathrm{foot}}$ (solid line) and total heating rates $\braket{Q_{\mathrm{turb}}+Q_{\mathrm{L}}}_{\mathrm{foot}}$ (dashed line) averaged over the coronal footpoint region ($6 \ \mathrm{Mm}\leq s \leq 14 \ \mathrm{Mm}$).}
  \label{fig:compare btw0050_0400}
\end{figure}

As described in Sec.~\ref{sec:parameter survey}, condensation becomes more likely to occur as the heating duration increases, the coronal magnetic field weakens, and the injected energy increases.
The dependence on the injected energy has already been discussed in previous studies.
Under steady localized heating, a criterion for the onset of condensation has been derived in terms of the ratio of the heating rates between the loop apex, $Q_{\mathrm{min}}$ and the loop footpoints, $Q_{\mathrm{foot}}$ \citep{klimchuk2019role}:
\begin{equation}
  \frac{Q_{\mathrm{foot}}}{Q_{\mathrm{min}}} > 1+\frac{c_{1}}{\Gamma_{\lambda_{\mathrm{H}}}}.\label{eq:KL2019}
\end{equation}
Here, $c_{1}\equiv R_{\mathrm{TR}}/R_{\mathrm{cor}}$ represents the ratio of the radiative losses per unit area integrated over the transition region to the coronal portion, respectively.
$\Gamma_{\lambda_\mathrm{H}}\equiv A_{\lambda_{\mathrm{H}}}/ A_{\mathrm{TR}}$ denotes the ratio of cross-sectional areas at a heating scale height above the transition region to that of the transition region.

This criterion was extended to the case of impulsive heating events by expressing it in terms of the ratio of the injected heating amount \citep{yoshihisa2025conditions}:
\begin{equation}
    \frac{\langle Q_{\mathrm{turb}}\rangle }{\langle Q_{\mathrm{L}} \rangle} \gtrsim 1+\frac{c}{\Gamma_{\lambda_{\mathrm{H}}}}.
\label{eq:KL2019_extended}
\end{equation}
Here, $\langle X \rangle$ denotes the spatiotemporal average of $X$ over
$s_{\mathrm{tr}} \le s \le s_{\mathrm{apex}}$
and
$t_{\mathrm{ini}} \le t \le t_{\mathrm{cool}}$,
defined as
\begin{equation}
    \braket{X}
    \equiv
    \frac{
    \int_{s_{\mathrm{tr}}}^{s_{\mathrm{apex}}} ds
    \int_{t_{\mathrm{ini}}}^{t_{\mathrm{cool}}} dt \, X
    }{
    \int_{s_{\mathrm{tr}}}^{s_{\mathrm{apex}}} ds
    \int_{t_{\mathrm{ini}}}^{t_{\mathrm{cool}}} dt
    }.    
\end{equation}
$s_{\mathrm{tr}}$ and $s_{\mathrm{apex}}$ is the location of transition region and loop apex.
$t_{\mathrm{ini}}$ is the start time of the localized heating, and $t_{\mathrm{cool}}$ is defined as $t_{\mathrm{ini}}+\tau_{\mathrm{rad}}$.
In \cite{yoshihisa2025conditions}, the heating duration $\tau_{\mathrm{w,eff}}$ was set to $\sim10^3 \ \mathrm{s}$.
Formula~\eqref{eq:KL2019_extended} does not take into account the effect of the heating duration or coronal background magnetic field on the condensation condition.

\subsubsection{Physical Interpretation of Heating Duration Dependence}\label{dis:heating duration}

To investigate the dependence on heating duration, Figure~\ref{fig:compare btw0050_0400} shows the temporal evolution of various physical quantities for two cases in Group B.
The injected energy is fixed at $E_{\mathrm{inj}}=0.75 \times 10^{11}\ \mathrm{erg\ cm^{-2}}$ in both cases.
The red curves represent the case with condensation ($\tau_{\mathrm{w,eff}} = 2000\ \mathrm{s}$), while the blue curves represent the case without condensation ($\tau_{\mathrm{w,eff}} = 300\ \mathrm{s}$).
Here, $\langle \cdots \rangle_{\mathrm{cor}}$ denotes the coronal average over the region extending 10 Mm above the transition region.
The transition region is defined by a temperature of $2 \times 10^{4}\ \mathrm{K}$.
The Field number $F_{\mathrm{i}}$ is defined as \citep{field1965thermal,yoshihisa2025conditions}:
\begin{equation}\label{Field number}
    F_{\mathrm{i}} = \frac{\tau_{\mathrm{cond}}}{\tau_{\mathrm{rad}}} = \frac{L}{2}\sqrt{\frac{n^{2}\Lambda(T)}{4\pi^{2} \kappa_{0}T^{7/2}}},
\end{equation}
where $\tau_{\mathrm{cond}}$ is the timescale of thermal conduction, and $\kappa_{0}=10^{-6}\ \mathrm{erg \ cm^{-1} \ s^{-1}\ K^{-7/2}}$ is the Spitzer conductivity \citep{spitzer1953transport}.
The dimensionless parameter $F_{\mathrm{i}}$ indicates thermal instability when the conductive timescale exceeds the radiative cooling timescale ($\tau_{\mathrm{cond}}>\tau_{\mathrm{rad}}$), i.e., when $F_{\mathrm{i}}>1$.

When the averaged total heating rate, $\braket{Q_{\mathrm{turb}} + Q_{\mathrm{L}}}_{\mathrm{foot}}$, exceeds the radiative cooling rate, $\braket{Q_{\mathrm{rad}}}_{\mathrm{foot}}$, (Figure~\ref{fig:compare btw0050_0400}(e)), the density increases (Figure~\ref{fig:compare btw0050_0400}(a)).
During this phase, thermal conduction transports energy from the localized heating region to the chromosphere. The resulting chromospheric heating increases the local thermal pressure, and the associated pressure gradient drives chromospheric evaporation.

After localized heating ceases, the evaporated plasma loses pressure support and transitions to gravitational draining, leading to a decrease in density. 
For longer heating durations, the onset of this draining is delayed, and the high-density state is maintained for a longer time (Figure~\ref{fig:compare btw0050_0400}(a)).
As a result, the radiative cooling rate, $n^{2}\Lambda(T)$, increases (Figure~\ref{fig:compare btw0050_0400}(d)), leading to more rapid temperature decrease and faster increase in the Field number (Figures~\ref{fig:compare btw0050_0400}(b, c)).
In the $\tau_{\mathrm{w,eff}} = 2000\ \mathrm{s}$ case, the condition $F_{\mathrm{i}} \gtrsim 1$ is satisfied while $Q_{\mathrm{rad}} \gg Q_{\mathrm{turb}}$, leading to the onset of condensation.
In the $\tau_{\mathrm{w,eff}} = 300\ \mathrm{s}$ case, the radiative cooling rate becomes comparable to the heating rate ($Q_{\mathrm{rad}} \sim Q_{\mathrm{turb}}$) before $F_{\mathrm{i}}$ exceeds unity, preventing condensation.

As discussed in Sec.~\ref{sec:parameter survey}, the critical injected energy required for condensation $E_{\mathrm{inj,crit}}$ increases as the heating duration $\tau_{\mathrm{w,eff}}$ decreases and becomes nearly saturated around $\tau_{\mathrm{w,eff}} \sim 150\ \mathrm{s}$.
This behavior can be understood as follows.
For shorter heating durations, the plasma begins to drain earlier.
To trigger condensation before the evaporated plasma drains completely, a larger amount of plasma must be supplied to the corona, requiring a larger injected energy.
When $\tau_{\mathrm{w,eff}} \lesssim 150 \ \mathrm{s}$, the heating duration becomes much shorter than the radiative cooling timescale ($\tau_{\mathrm{w,eff}} \ll \tau_{\mathrm{rad}}$).
In this regime, further reductions in $\tau_{\mathrm{w,eff}}$ do not significantly change the cooling process.
As a result, the required injected energy saturates.

Our parameter survey shows that $E_{\mathrm{inj,crit}}$ increases from the steady-heating value to at most about an order of magnitude larger as the heating duration becomes shorter. 
This suggests that Equation~\eqref{eq:KL2019_extended} can be generalized as
\begin{equation}
    \frac{\langle Q_{\mathrm{turb}} \rangle}{\langle Q_{\mathrm{L}} \rangle} \gtrsim f(\tau_{\mathrm{w}})\Bigl(1+\frac{c_{1}}{\Gamma_{\lambda_{\mathrm{H}}}}\Bigr),
\label{eq:KL2019_final}
\end{equation}
where $f(\tau_{\mathrm{w}})$ increases from $1$ to $\lesssim 10$ as the heating duration $\tau_{\mathrm{w,eff}}$ decreases.

\subsubsection{Physical Interpretation of Magnetic Field Dependence}\label{dis:magnetic field}

\begin{figure}[!]
  \epsscale{1.0}
  \plotone{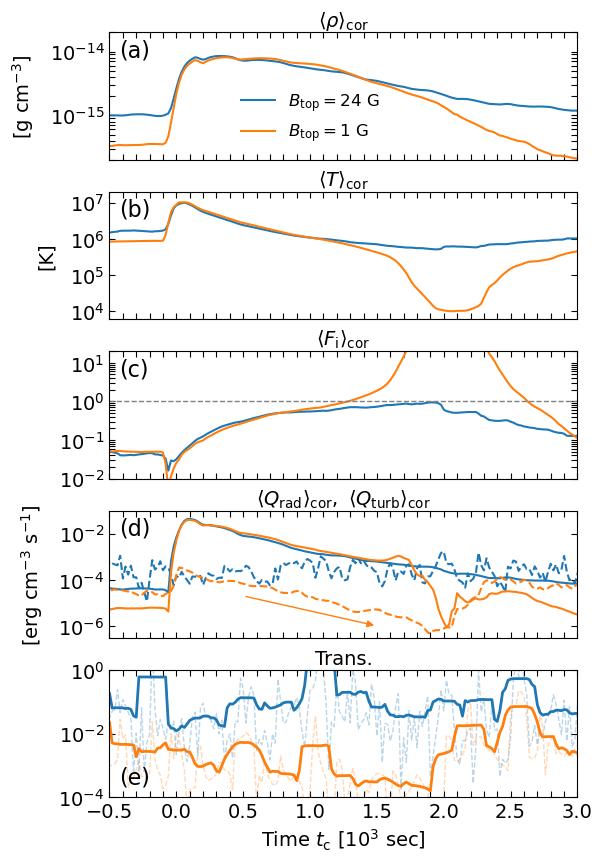}
  \caption{Time evolutions of various physical quantities for two cases in Group C with the same injected energy, $E_{\mathrm{inj}}=0.94 \times 10^{11} \ \mathrm{erg \ cm^{-2}}$, are shown.
  The blue and orange lines indicate the cases without ($B_{s,\mathrm{cor}}=24 \ \mathrm{G}$) and with ($B_{s,\mathrm{cor}}=1 \ \mathrm{G}$) condensation, respectively.
  Panels (a)--(d) show the same quantities as in Figure~\ref{fig:compare btw0050_0400}.
  Panel (e) shows the transmittance ratio of Poynting flux from near the surface ($s=2\ \mathrm{Mm}$) to the corona ($s=30\ \mathrm{Mm}$).
  Dashed lines with lighter colors represent the raw data, while solid lines show values smoothed over 200 s.}
  \label{fig:compare fm0041_1000}
\end{figure}

Figure~\ref{fig:compare fm0041_1000} shows the temporal evolution of two cases in Group C with the same injected energy ($E_{\mathrm{inj}} = 0.94\ \mathrm{erg\ cm^{-2}}$).
As shown in Section~\ref{sec:Coronal Loops}, the initial coronal conditions differ between these cases.
Despite this difference, their temporal evolution remains similar until $t_{\mathrm{c}} \sim 1100\ \mathrm{s}$ after the onset of localized heating.
This is because the imposed localized heating is much stronger than the background heating, producing a similar thermodynamic response in both cases.

The evolution begins to diverge after $t_{\mathrm{c}} \sim 1100\ \mathrm{s}$.
In the stronger-field case, the radiative cooling rate becomes comparable to the turbulent heating rate at $t_{\mathrm{c}} \sim 1400\ \mathrm{s}$ (Figure~\ref{fig:compare fm0041_1000}(d)).
Subsequently, the cooling process---characterized by a temperature decrease and an increase in the Field number $F_{\mathrm{i}}$---ceases, and the system gradually returns to its initial state.
In the weaker-field case, the condition $Q_{\mathrm{rad}} \gg Q_{\mathrm{turb}}$ is maintained beyond $t_{\mathrm{c}} \sim 1100\ \mathrm{s}$, and $F_{\mathrm{i}} \gtrsim 1$ is satisfied, leading to condensation.
This difference can be attributed to two factors.
The first is the lower initial turbulent heating rate $Q_{\mathrm{turb}}$ in the weaker-field case.
The second is the substantial reduction of $Q_{\mathrm{turb}}$ during the cooling phase, indicated by the orange arrow in Figure~\ref{fig:compare fm0041_1000}(d).

The first reason follows directly from the criteria for onset of condensation (formula~\eqref{eq:KL2019} or \eqref{eq:KL2019_final}), which indicate that condensation is more likely to occur under weaker background heating conditions corresponding to weaker magnetic fields.
Figure~\ref{fig:parameter survey}(c) shows that the critical injected energy required for condensation scales as
\begin{equation}
    E_{\mathrm{inj,crit}} \propto B_{s,\mathrm{cor}}^{0.26} \label{eq:E_inj_crit vs Bs}
\end{equation}
for $2 \leq B_{s,\mathrm{cor}} \leq 100\ \mathrm{G}$.
Similarly, Figure~\ref{fig:Bs_vs_roteQtbeta}(c) shows that the background heating rate per unit mass scales as
\begin{equation}
    \frac{Q_{\mathrm{turb}}}{\rho} \propto B_{s,\mathrm{cor}}^{0.31}
\end{equation}
for $1 \leq B_{s,\mathrm{cor}} \leq 100\ \mathrm{G}$.
The similarity between these scalings suggests that the magnetic-field dependence of the condensation threshold is primarily controlled by the background turbulent heating rate.

The second reason is the reduction of the turbulent heating rate during the cooling process in the weaker-field case.
As coronal rain forms in the lower corona, the resulting density enhancement produces a strong gradient in the Alfv\'en speed,
$ B_s/\sqrt{4\pi\rho}$, which enhances the reflection of Alfv\'en waves and reduces the upward Poynting flux (Figure~\ref{fig:compare fm0041_1000}(e)).
The reduced Poynting flux subsequently suppresses the turbulent heating rate in the corona, thereby further facilitating condensation.
This additional suppression of turbulent heating causes the critical injected energy $E_{\mathrm{inj,crit}}$ in the
$B_{s,\mathrm{cor}} = 1\ \mathrm{G}$
case to deviate significantly from the scaling relation and fall well below the expected trend (Figure~\ref{fig:parameter survey}(c)).
Although similar behavior is also observed for
$B_{s,\mathrm{cor}} = 8\ \mathrm{G}$,
the deviation becomes increasingly pronounced toward weaker magnetic fields.


\subsection{The Origin of Coronal Rain near Footpoint}\label{sec:Discussion Morphology}

\begin{figure}[!]
  \epsscale{1.0}
  \plotone{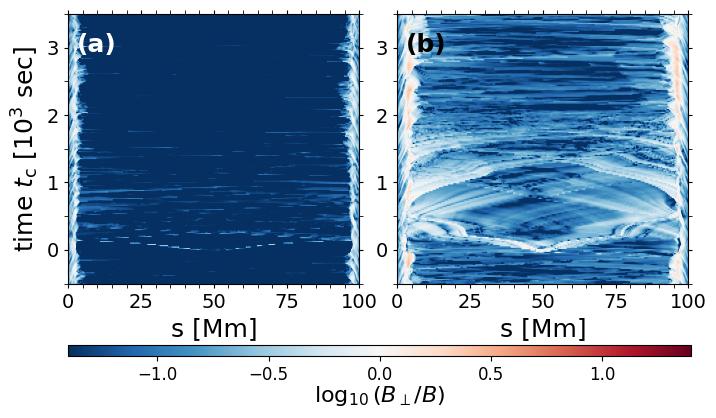}
  \caption{Space-–time diagrams of $\log_{10}(B_{\perp} / B_s)$ for (a) $B_{s,\mathrm{cor}}=24\ \mathrm{G}$ and (b) $B_{s,\mathrm{cor}}= 4\ \mathrm{G}$ in Group C, where $B_{\perp} = \sqrt{B_y^2 + B_z^2}$ is the amplitude of transverse magnetic field perturbation. 
  This quantity represents the degree of Alfv\'enic nonlinearity.}
  \label{fig:compare dB_B}
\end{figure}

The morphology of coronal rain depends on the coronal magnetic field strength (Section~\ref{sec:Dependence on Magnetic Field Strength}).
For $B_{s,\mathrm{cor}} \lesssim 8\ \mathrm{G}$, multiple cool plasma condensations form at the lower corona. 

To investigate the origin of this behavior, we examine the nonlinearity of Alfv\'en waves.
Figure~\ref{fig:compare dB_B} shows space-–time diagrams of $\log_{10}(B_{\perp} / B_s)$ for $B_s = 24\ \mathrm{G}$ (left) and $4\ \mathrm{G}$ (right), where $B_{\perp} = \sqrt{B_x^2 + B_y^2}$ represents the amplitude of transverse magnetic field perturbation. 
The level of nonlinearity in the quasi-steady state is higher in the weaker-field case than in the stronger-field case.
In both cases, the nonlinearity is further enhanced when the coronal density increases after the localized heating.
For $B_{s,\mathrm{cor}} = 24\ \mathrm{G}$, the average of $\log_{10}(B_{\perp}/B_{s})$ increases from $\sim 0.015$ (maximum $\sim 0.08$) before heating to $\sim 0.03$ (maximum $0.15$) after heating ($500 \ \mathrm{s} \leq t_{\mathrm{c}} \leq 1500 \ \mathrm{s}$).
For $B_{s,\mathrm{cor}} = 4\ \mathrm{G}$, the increase is more pronounced, from $\sim 0.1$ (maximum $\sim 0.3$) to $\sim 0.17$ (maximum $0.7$).
This indicates that the weaker-field case reaches a higher level of nonlinearity.
Such an enhanced nonlinearity suggests more efficient mode conversion of Alfv\'en waves into compressive modes in the weaker-field case, leading to stronger density fluctuations.
These density fluctuations are likely responsible for the formation of multiple blob-like structures near the footpoints.

The dependence of the nonlinearity on $B_{s,\mathrm{cor}}$ can be qualitatively understood as follows.
We consider a quasi-steady corona before the localized heating, or during the gradual density decrease after the localized heating, and assume $\rho = \mathrm{const.}$, $v_{s} \approx 0$, and $B_{s} = \mathrm{const.}$
Under these assumptions, linearizing the momentum and induction equations (Equation~\eqref{vperp}, \eqref{Bperp}) with perturbations of the form $v_{\perp}, B_{\perp} \propto \exp\left[i(ks - \omega t)\right]$ yields
\begin{equation}
B_{\perp} = -\frac{k}{\omega} B_{s} v_{\perp} = \mp \sqrt{4\pi \rho}\, v_{\perp}.
\end{equation}
Using the conservation of Poynting flux,
\begin{equation}
\begin{aligned}
\frac{1}{4}\rho \left(\xi_x^2 + \xi_y^2\right) v_A
&\sim \frac{1}{2} \rho^{1/2} \frac{B_{\perp}^2}{4\pi \rho} B_{s} \\
&\sim \rho^{-1/2} B_{\perp}^{2} B_{s} \sim \mathrm{const.},
\label{eq:poynting flux conversion}
\end{aligned}
\end{equation}
which implies that $B_{\perp}$ increases with increasing density.

Finally, the stronger nonlinearity in the weaker-field case can also be understood from the dependence on $B_s$.
From Figure~\ref{fig:parameter survey}, we obtain $\rho \propto B_{s,\mathrm{cor}}^{0.32}$.
Substituting this relation into the above expression, we find that the nonlinearity exhibits a power-law dependence on the magnetic field strength,
\begin{equation}
\frac{B_{\perp}}{B_{s}}
\propto B_{s}^{-1.4}.
\end{equation}
This scaling indicates that the nonlinearity becomes stronger as the magnetic field strength decreases.

\subsection{Comparison with Previous Studies}

Previous studies assuming quasi-steady footpoint-concentrated heating typically produce condensations at a specific position along the loop, resulting in coronal rain that falls toward one footpoint \citep[e.g.,][]{muller2003dynamics,muller2004dynamics,antolin2010coronal,mikic2013importance,fang2015coronal,froment2017long,froment2018occurrence,kohutova2020self,antolin2022multi,pelouze2022role,daley2024simulating,lu2024periodic}.
In contrast, our low-frequency impulsive heating model leads to cooling throughout the coronal portion of the loop, producing more spatially extended condensations that subsequently drain toward both footpoints (Section~\ref{sec:representative case}).
This occurs because the localized heating duration is much shorter than the radiative cooling timescale, allowing mass redistribution along the loop, which results in a nearly uniform coronal density.
As a result, thermal instability develops nearly simultaneously throughout the coronal loop.

In Section~\ref{sec:Dependence on Magnetic Field Strength}, we showed that the morphology of coronal rain depends on the coronal magnetic field strength.
In weaker-field cases, the nonlinearity of Alfvén waves becomes significantly enhanced, leading to the formation of fragmented condensations (Section~\ref{sec:Discussion Morphology}).
A similar tendency has been reported by \cite{xia2017coronal}, who performed 3D simulations of coronal rain formation in a weak coronal magnetic field environment ($\sim3$ G).
They found that condensations formed near the apex of a bipolar loop subsequently fragmented into smaller blobs through the development of the Rayleigh–Taylor instability.
They further suggested that such fragmentation may be suppressed in stronger magnetic fields where the instability growth is inhibited.
Although the fragmentation mechanism identified in their study differs from ours, as it relies on multidimensional effects absent from our 1.5D model, both studies suggest that weak coronal magnetic fields favor the formation of fragmented coronal rain.

Several previous numerical studies have also investigated condensation triggered by low-frequency impulsive heating.
Early work focused on post-flare loops, where the corona was found to cool uniformly without forming blob-like structures \citep{reep2020electron}.
\cite{benavitz2025spatiotemporal} extended this framework by incorporating spatial and temporal variations in elemental abundances due to the first ionization potential (FIP) effect.
They showed that low-FIP elements are transported toward the loop apex following chromospheric evaporation, locally enhancing radiative cooling and leading to localized condensation.
Our simulations, on the other hand, showed the formation of multiple blob-like coronal rain condensations.
This behavior is a distinct feature associated with MHD waves driven by velocity perturbations at the footpoints.

\cite{ruan2024lorentz} performed a 3D MHD simulation covering the impulsive and gradual phases of a solar flare.
They found that the post-flare loop system underwent widespread cooling due to thermal conduction and radiative losses.
Coronal rain formed through catastrophic cooling, with its formation progressing from the inner to the outer loops.
The subsequent drainage of the coronal rain is qualitatively similar to that found in our cases with $\tau_{\mathrm{w,eff}}<\tau_{\mathrm{rad}}$, suggesting that the effective energy input duration of an individual loop was shorter than its cooling timescale.

Condensation driven by a single impulsive heating event has also been discussed in the context of prominence formation \citep{huang2021unified,huang2025unified, yoshihisa2025conditions}.
\cite{yoshihisa2025conditions} examined condensation along dipped magnetic field lines under impulsive heating similar to that considered here and found that multiple condensations initially form and subsequently merge into a single prominence.
This suggests that the early evolution of cooling plasma is qualitatively similar across different loop geometries.

Observational studies also provide important constraints on the heating timescale responsible for coronal rain formation.
\cite{kohutova2019formation} reported coronal rain associated with a single impulsive nanoflare heating event.
Their Figure~6 shows, in an AIA 304~\AA\ time–distance diagram, coronal rain forming and draining toward both loop footpoints.
Such behavior is consistent with the scenario proposed in this study, in which low-frequency impulsive heating triggers loop-wide condensation and drainage toward both footpoints.
Similar pattern is observed in coronal loops exhibiting periodic TNE cycles, where cooling plasma appears nearly simultaneously along the entire coronal portion of the loop and falls toward both footpoints \citep{froment2020multi}.
Although TNE cycles are generally attributed to quasi-steady footpoint-concentrated heating, our results raise the possibility that low-frequency impulsive heating events embedded within such a heating environment trigger the condensation and contribute to the observed loop-wide cooling and drainage.
In contrast, \cite{antolin2012observing} reported coronal rain that forms near the loop apex and subsequently drains downward.
This behavior resembles the regime identified in our simulations where the heating duration exceeds the radiative cooling timescale ($\tau_{\mathrm{w,eff}} > \tau_{\mathrm{rad}}$), leading to localized condensation.

These results suggest that the spatial distribution and drainage pattern of coronal rain may provide diagnostics of the coronal heating timescale.
The loop-wide cooling followed by drainage toward both footpoints may indicate low-frequency impulsive heating ($\tau_{\mathrm{w,eff}} < \tau_{\mathrm{rad}}$), whereas localized condensations forming at a specific position may be indicative of quasi-steady heating ($\tau_{\mathrm{w,eff}} > \tau_{\mathrm{rad}}$).
Therefore, coronal rain dynamics can serve as an observational probe of coronal heating timescales.

\cite{karpen2005prominence} showed that variations in the flux tube cross-sectional area within the coronal part of the loop can directly modify field-aligned flows and condensation dynamics.
The flux tube expansion in our model is largely confined to the lower atmosphere, while the cross-sectional area is nearly uniform in the coronal part of the loop.
Thus, the direct geometrical effect of the coronal cross-sectional variation found by \cite{karpen2005prominence} is expected to be limited in our simulations.
Nevertheless, varying $B_{s,\mathrm{cor}}$ through the expansion factor can indirectly affect Alfv\'en wave reflection and transmission.
An alternative setup would be to keep the expansion factor fixed and vary the footpoint magnetic field strength $B_{s,\mathrm{surf}}$,
thereby varying $B_{s,\mathrm{cor}}$ consistently.
This setup would also alter the Alfv\'en speed in the lower atmosphere and the Poynting flux injected through the lower boundaries.
Thus, neither parameterization completely isolates the effect of $B_{s,\mathrm{cor}}$.
The Group C results should therefore be interpreted as the combined effects of $B_{s,\mathrm{cor}}$ and the associated lower atmospheric expansion, rather than as the pure effect of $B_{s,\mathrm{cor}}$.

An additional multidimensional effect may influence the dynamics of falling coronal rain.
\cite{hillier2025understanding} showed that coronal rain blobs can entrain surrounding coronal plasma during their descent.
This interaction increases the effective inertia of the condensations and leads to deceleration relative to free-fall motion.
Because our model is one-dimensional, such momentum exchange with the ambient corona is not captured.
Incorporating these multidimensional effects may be important for a more quantitative comparison with observed coronal rain velocities.

\subsection{Why is coronal rain preferentially observed in active regions?}
In the solar corona, coronal rain is predominantly observed in strong-field environments such as active regions (cf. \citealt{antolin2022multi}).
We found that the critical injected energy required for condensation, $E_{\mathrm{inj,crit}}$, follows the scaling relation given by Formula~\eqref{eq:E_inj_crit vs Bs}.
If the energy released by impulsive heating events scales with the magnetic energy, such that $E_{\mathrm{inj}} \propto B_{s,\mathrm{cor}}^{2}$, the increase in injected energy with magnetic field strength is substantially stronger than the increase in the condensation threshold.
Under this assumption, stronger-field loops may more readily satisfy the condensation criterion despite their higher critical energy threshold.
This can explain why coronal rain is preferentially observed in active-region loops.

\section{Conclusion}\label{sec:summary}

We performed 1.5D MHD simulations of plasma condensation in coronal loops triggered by a single impulsive heating event.
The model includes velocity perturbations and a phenomenological turbulent heating term, enabling more realistic coronal heating and loop dynamics.
Our previous study showed that condensation can occur even when the heating duration is shorter than the radiative cooling timescale ($\tau_{\mathrm{w,eff}} < \tau_{\mathrm{rad}}$), provided that sufficient energy is injected.

In this study, we systematically investigated the effects of the heating duration $\tau_{\mathrm{w,eff}}$ and the coronal magnetic field strength $B_{s,\mathrm{cor}}$ on coronal rain formation and dynamics.

For the dependence on heating duration, we performed simulations with one-sided localized heating (Group A) and both-sided localized heating (Group B).
In Group B, condensation occurs over the entire range $\tau_{\mathrm{w,eff}} = 3$–$3500\ \mathrm{s}$.
In Group A, condensation does not occur for $\tau_{\mathrm{w,eff}} > \tau_{\mathrm{rad}} \sim 2000\ \mathrm{s}$ because asymmetric heating drives a unidirectional flow that removes the plasma before condensation develops, consistent with previous studies \citep[e.g.,][]{mikic2013importance,klimchuk2019role}.
When $\tau_{\mathrm{w,eff}} < \tau_{\mathrm{rad}}$, coronal rain forms over a broad coronal region and drains toward both loop footpoints.
This behavior differs from the conventional steady heating scenario ($\tau_{\mathrm{w,eff}} > \tau_{\mathrm{rad}}$), where condensations form at specific locations typically fall toward a single footpoint.

The injected energy required for condensation also depends on the heating duration.
In both-sided heating cases, the required injected energy increases with decreasing heating duration and becomes nearly saturated for $\tau_{\mathrm{w,eff}} \lesssim 150$ s at a value about an order of magnitude larger than that for long-duration heating.
In one-sided heating cases, the required injected energy decreases again for very short heating durations because strong localized heating drives substantial evaporation from the opposite chromosphere.

We also investigated the dependence on the coronal magnetic field strength over the range $B_{s,\mathrm{cor}} = 1$--$100\ \mathrm{G}$.
Weaker magnetic fields produce lower background turbulent heating rates, making condensation easier to trigger.
The injected energy required for condensation approximately scales with the background turbulent heating rate.
The morphology of coronal rain also changes with magnetic field strength.
For $B_{s,\mathrm{cor}} \lesssim 8\ \mathrm{G}$, blob-like condensations become prominent near the loop footpoints.
This behavior is caused by enhanced nonlinearity of Alfv\'en waves under weaker magnetic fields.
The increased nonlinearity promotes more efficient mode conversion into compressive waves, generating density perturbations that trigger multiple blob-like condensations.
This tendency is particularly pronounced for $B_{s,\mathrm{cor}} = 1\ \mathrm{G}$, where coronal rain formed near the footpoints reflects upward-propagating Alfv\'en waves.  
This reduces the upward Poynting flux, leading to a decrease in the turbulent heating rate and making condensation more likely.

Direct observational determination of coronal heating properties and magnetic field strengths remains difficult.
Our results suggest that these quantities can instead be constrained indirectly through the dynamics and morphology of coronal rain.
Future work should test this framework by combining spectropolarimetric measurements of coronal magnetic fields using coronal rain \citep[e.g.,][]{schad2016vector,kriginsky2021magnetic} with statistical analyses of coronal rain morphology and dynamics.

\vskip\baselineskip

This work was supported by JST SPRING, Grant Number JPMJSP2110.
Numerical computations in this study were carried out on the PC cluster at the Center for Computational Astrophysics, National Astronomical Observatory of Japan, and at the Yukawa Institute Computer Facility in Kyoto University.
The authors are grateful to the anonymous referee for improving the manuscript.

\bibliographystyle{aasjournal}
\bibliography{reference_0421}
\end{document}